\documentclass[preprint,1p,times]{elsarticle}
\usepackage{array}
\usepackage{graphicx} 
\usepackage{amsmath}
\usepackage{amssymb}
\usepackage{graphicx}
\usepackage{lineno}
\usepackage{hyperref}
\usepackage{textcomp}
\usepackage{natbib}
\usepackage{microtype}   
\journal{Nuclear Instruments and Methods in Physics Research Section A}
\usepackage{stfloats}
\begin{document}

\begin{frontmatter}

\title{R\&D Efforts in Cherenkov Imaging Technologies for Particle Identification in Future Experiments}

\author{Chandradoy Chatterjee\corref{corresponding author}}
\ead{chandradoy.chatterjee@ts.infn.it}
\address{Istituto Nazionale di Fisica Nucleare (INFN), Sezione di Trieste, Italy}
\begin{abstract}
Cherenkov imaging detectors will continue to play a central role for particle identification in future particle and nuclear physics experiments. Growing demands on momentum coverage, timing precision, radiation tolerance, and sustainability have driven extensive R\&D in detector concepts, radiator materials, and photon sensors. This article reviews recent efforts, focusing on experiments leading advances in sensor technology, radiator materials, and the exploitation of Cherenkov photon timing to push PID limits, while highlighting synergies across experiments in addressing common challenges.
\end{abstract}

\begin{keyword}
Cherenkov detectors \sep Particle identification \sep RICH \sep DIRC \sep MCP-PMT \sep SiPM
\end{keyword}

\end{frontmatter}
\section{Introduction}
Future upgrades of the ALICE \cite{ALICE} and LHCb \cite{LHCb_performance} experiments at the CERN LHC, the ePIC  experiment at the Electron–Ion Collider (EIC) \cite{EIC_YR} at BNL, USA, and the PANDA \cite{PANDA_TDR} and CBM \cite{CBM} experiments at FAIR are among the main drivers of R\&D efforts in Cherenkov imaging technologies. However, the development of these technologies is not limited to these projects alone. EIC-China \cite{EIC_China}, upgrade of the Belle II \cite{BelleII}, ALADDIN \cite{ALADDIN_2} at CERN and the CLD detector at the CERN FCC \cite{CLD} are also making significant contributions to the advancement of Cherenkov imaging technologies. These leading future experiments share common challenges and solutions, which are being addressed within the newly formed DRD4 collaboration for detector R\&D on particle identification (PID) and photon detectors \cite{DRD4}. DRD4 plays a central role in coordinating collaboration, fostering innovation, and advancing Cherenkov imaging technology. In this article, we review selected future experiments and upgrades of existing detectors that make extensive use of PID based on Cherenkov imaging; we will look into their physics objectives, the challenges imposed on PID technologies, the corresponding risk-mitigation strategies, and the associated R\&D efforts, with particular emphasis on sensor technologies and optimization of radiator materials.

\section{Use of PID in upgraded LHC experiments:}
\subsection{ALICE-3 upgrade}
\label{sec:ALICE3}
The proposed upgrade of the ALICE detector at CERN aims to realize a completely new experimental apparatus, ALICE3, for LHC Run 5 and beyond. This upgrade seeks to study the properties of the quark–gluon plasma in relativistic heavy-ion collisions with a level of precision and accuracy exceeding current limits. Extensive PID capabilities are central to the ALICE3 upgrade. For example, the design target of the ALICE3 PID system is to achieve better than 3$\sigma$ e/$\pi$, $\pi$/K and K/p separation for momenta up to 2~GeV/c, 10~GeV/c, and 16~GeV/c, respectively, while providing acceptance over eight units of pseudorapidity ($|\eta|\leq 4$). The nature of collider experiments limits the available space, and the high LHC luminosity results in a harshly crowded environment~\cite{ALICE_LOI}.
\par
To meet these PID requirements, the ALICE3 baseline design foresees the use of ultra-fast time-of-flight (TOF) and proximity focusing RICH detectors in both the barrel region ($|\eta|\leq 2$), referred to as bRICH, and the endcap region ($2\leq|\eta|\leq 4$), referred to as fRICH. Silicon photomultipliers (SiPMs) have been chosen as the baseline photo-sensors for the barrel region due to the presence of a solenoidal magnetic field of approximately 2~T. However, the choice of photo-sensors in the endcap region is more critical and remains subject to ongoing investigation and further research. The radiation level in the endcap regions is approximately two orders of magnitude higher (3.9$\times10^{13}$ MeV n$_{eq}$/cm$^2$) than in the barrel region (6.1$\times10^{11}$ MeV n$_{eq}$/cm$^2$). It has been observed that SiPM performance suffers severe limitations as the dark count rate (DCR) increases to tens of MHz/mm$^2$ when the radiation level reaches $10^{13}$~MeV n$_{eq}$/cm$^2$ \cite{ALICE_Rad1,ALICE_SiPM}. Commercial atomic-layer-deposition (ALD) coated MCP-PMTs have been reported to operate in high radiation levels without noticeable degradation; however, their performance has been noted to degrade above about 20~Ccm$^{-2}$ \cite{ALICE_Lehmann_MCP}. In addition, Large-Area Picosecond Photodetectors (LAPPDs) and High-Rate Picosecond Photodetectors (HRPPDs) are also under consideration. These efforts highlight the synergy between photo-sensor development for the ePIC experiment and the ALICE3 upgrade. Presently, the choice of optimal photo-sensors for the endcap RICH remains an open question due to the level of radiation. Further studies and R\&D efforts would be essential to identify the photo-sensors for the endcap RICH. 
\begin{figure*}
    \centering
    \includegraphics[width=0.62\textwidth]{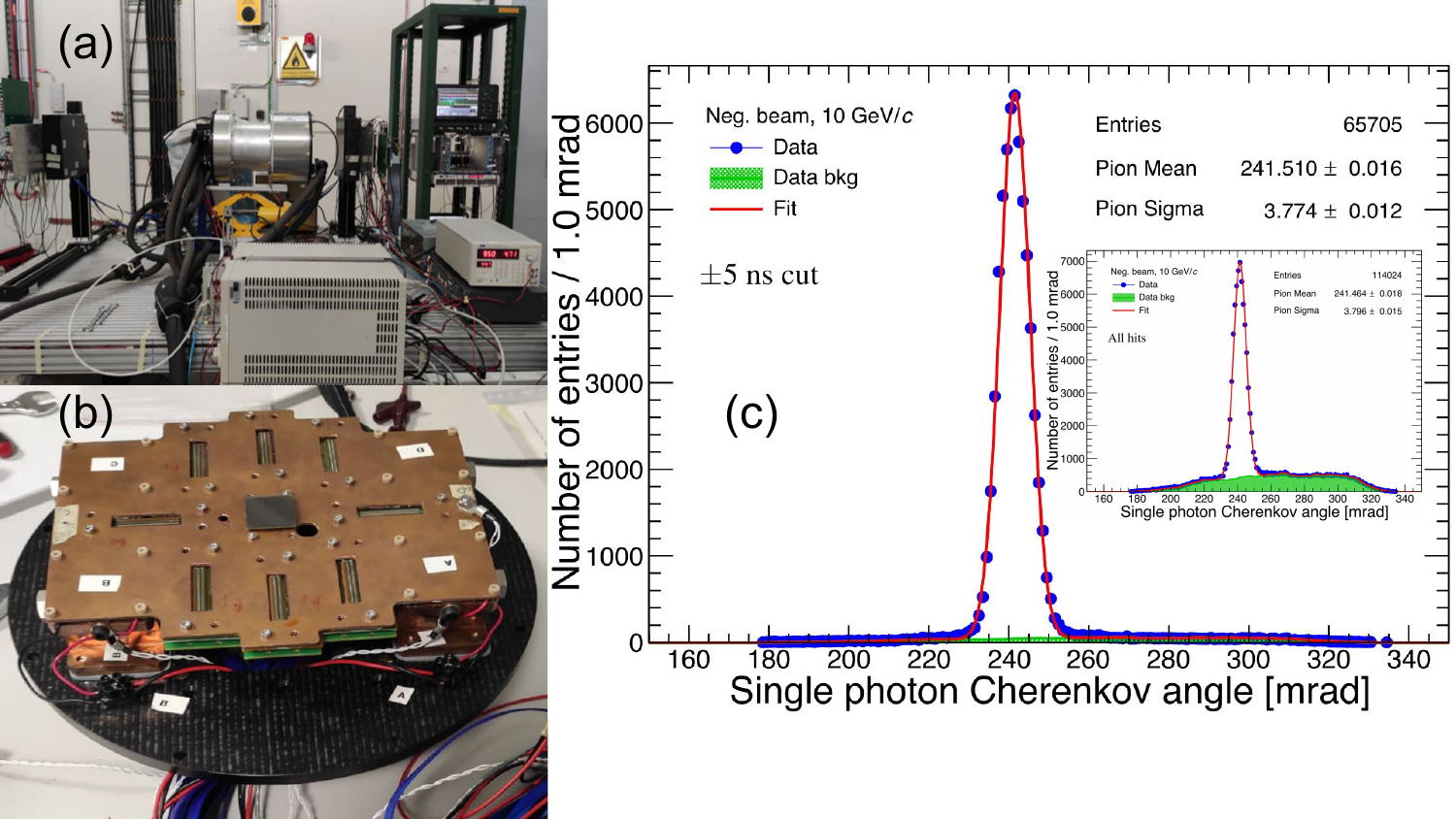}
    \vskip-1.2em
    \caption{Highlights of beam test with SiPM based proximity focusing RICH for ALICE3 upgrade: (a) and (b) Setup for the beam test measurements, (c) Cherenkov angle distribution; with a 5~ns time cut a significant improvement is seen in signal to noise ratio. The level of background in Cherenkov angle distribution without any timing filter is shown in the inner panel.}
    \label{fig:ALICE}
\end{figure*}
The use of SiPMs operated as TOF detectors has also been considered within the ALICE3 upgrade, motivated by promising R\&D results \cite{ALICE_Nicassio}. A thin layer of high-refractive-index material (SiO$_2$, MgF$_2$, or high-refractive-index glass) placed directly in front of the SiPM array allows the production of a large number of Cherenkov photons. Owing to the 1/$\sqrt{N}$ dependence of the time resolution, where $N$ is the number of detected photons, a timing resolution of approximately 20~ps is expected when five or more photons are detected. Given the excellent intrinsic performance of SiPMs, such performance is achievable and has been demonstrated in recent beam test studies for the ALICE3 upgrade \cite{ALICE_Nicassio}. Furthermore, the use of multiple aerogel layers with different refractive indices as a radiator is under consideration to minimize emission point uncertainty and increase the number of detected photons. It is also being considered to fill the proximity gap with a C$_5$F$_{10}$O/N$_2$ gas mixture, enabling operation of the bRICH as a threshold counter to veto electrons up to momenta of approximately 4~GeV/c. This option would require increasing the proximity gap between the radiator and the sensor from its nominal value of 20~cm to 35~cm. Further details can be found in \cite{ALICE_Scope}.

A dedicated beam test was performed to validate the design principles of the bRICH at the CERN PS T10 beam test facility. The beam test used a setup with a hydrophobic aerogel tile as the radiator, produced by the Aerogel Factory Co. Ltd. (Chiba, Japan) \cite{ALICE_ADACHI}, with a refractive index of 1.03. This choice of aerogel radiator is of synergistic interest with the ePIC experiment at the EIC. The setup was designed using two different types of SiPM sensors produced by Hamamatsu. Sensors from the S13361 series with 64 pixels were used to detect Cherenkov photons generated in a 1~mm thick quartz window acting as a secondary Cherenkov radiator, while sensors from the S13552 series with 128 channels were used to detect Cherenkov photons produced in the aerogel. In Fig.~\ref{fig:ALICE} (a) and Fig.~\ref{fig:ALICE} (b) we can see the set-up for their beam test. In Fig.~\ref{fig:ALICE} (c), we can also notice the impact of 5~ns time gating. The results are promising and are in excellent agreement with expectations: (1) a Cherenkov angle of 242~mrad for saturated particles; (2) a single-photon angular resolution of 3.8~mrad; and (3) scaling with the number of detected photons allows the required and targeted bRICH angular resolution of 1.5~mrad to be achieved, along with substantial suppression of background hits when an appropriate timing window is applied. It is expected to couple the final ASIC design to annealed SiPMs operated at $-40^\circ$C after exposure to the radiation levels anticipated in the bRICH region~\cite{ALICE_TB}. 

\subsection{LHCb upgrade}
\label{sec:LHCb}
\begin{figure*}[!thb]
   \centering
        \includegraphics[width=0.62\textwidth]{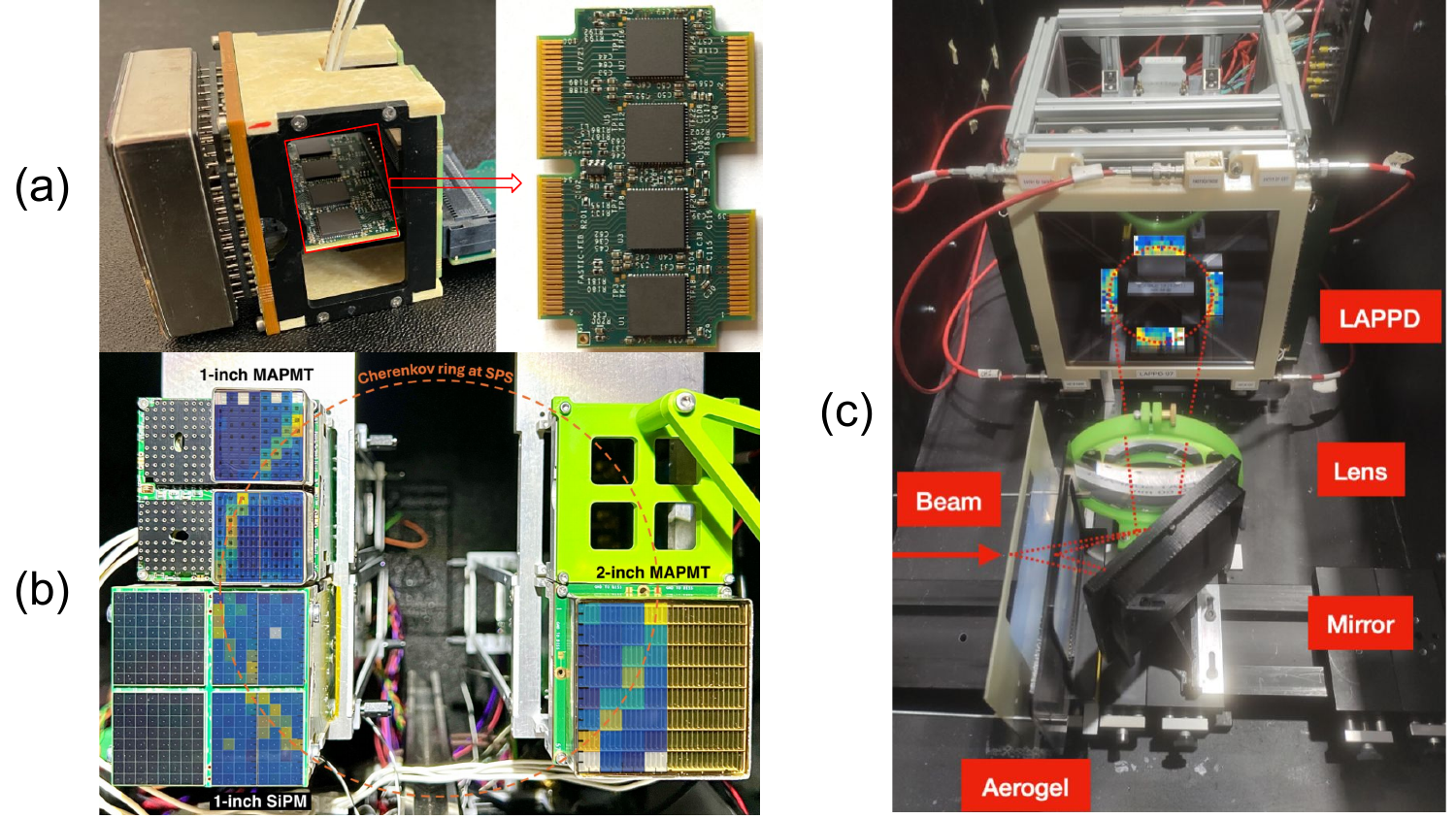}
        \vskip-1.2em
        \caption{Highlights of R\&D activities related to PID systems of LHCb: (a) coupling of new ASIC into photo-sensors, (b) beam test setup with standard MaPMT and SiPM, and (c) beam test setup using LAPPD.}
        \label{fig:LHCb_RICH_all}    
\end{figure*}

The PID system of the LHCb \cite{LHCB_RICH_TDR} experiment currently comprises two RICH detectors to cover an unprecedented phase-space. RICH1 \cite{LHCb_RICH1} is placed upstream of the magnet, and another named RICH2 \cite{LHCb_RICH2} is placed downstream of the spectrometer magnet. The radiator gas in RICH1 has a Cherenkov threshold of $\sim$10~GeV/c for kaons and charged hadron PID is possible down to $\sim$ 2~GeV/c in the veto mode. The radiator gas in RICH2 provides good $\pi$/K separation in the high momentum particle range up to $\sim$100~GeV/c \cite{LHCb_performance}. The aim of the future upgrade of the LHCb PID sub-systems is to provide similar good performance in the HL-LHC era \cite{LHCb_Upgrade_TDR}. This would require a substantial improvement in both the precision of the measurement of the Cherenkov angle and the rejection of the background. Both the improvements can be achieved with larger number of detected photons. 
The advanced RICH reconstruction algorithm shows that the difference between the measured photon Time-of-Arrival (TOA) and the prediction from full LHCb detector simulation implies an intrinsic RICH timing resolution below 10 ps. A 5 ns timing gate is sufficient to reject a significant fraction of the background. Currently, the photon detector chain for the RICH system consists of Multi-anode photomultiplier tubes (MaPMTs) read out by the CLARO ASICs on front-end board~\cite{LHCb_CLARO}. For running between 2030 and 2033, the RICH detectors are expected to replace the CLARO ASIC with a fast ASIC known as FastRICH \cite{LHCb_FastIC},\cite{LHCb_FastRICH} with no changes in the photo-sensors. A configurable time gate reduces the TDC range to a configurable width up to 6.25~ns (with 25~ps TDC bins) or up to 25~ns (100~ps bins); therefore, reducing the outlier hits and requiring fewer bits per hit timestamp. The requirements for the future RICH readout ASIC are several: reduction of the bandwidth of the outgoing data, power consumption and radiation hardness, in addition to timing performance. The Upgrade II foresees the ASIC to withstand a fluence of approximately 2x10$^{13}$ n$_{eq}$/cm$^2$ and 12 kGy at the RICH1 sensor column, including a factor two safety margin on the predicted fluence. Furthermore, the ASIC should be compatible to lpGBT~\cite{lpGBT} and VL+ \cite{VLPlus} radiation-hard chipset used for the LHCb optical readout. 
In Fig.~\ref{fig:LHCb_RICH_all} (a), we see the ASIC coupled to the classical MaPMT. The ASIC is versatile and it can be coupled to various other types of photo-sensors, namely SiPMs, and MCP-PMT-based LAPPDs. This brings us to understand the second part of the LHCb upgrade plan, namely during 2036 and beyond. On one hand, there is available time to carry on R\&D activities to identify the optimal photo-sensor; on the other hand with the increasing challenges to keep the single photon resolution as good as it is now. Therefore, the quest is for a photocathode with high photon detection efficiency shifted towards larger wavelength. This allows minimizing the chromatic error and ensuring good single photon angular resolution. Beam tests are conducted with classical MaPMTs, and SiPMs as well as LAPPD like photo-sensors coupled to FastRICH ASICs to identify the optimal choice of photo-sensors (Fig.~\ref{fig:LHCb_RICH_all} (b) and Fig.~\ref{fig:LHCb_RICH_all} (c)). SiPM coupled to a micro-lens array are envisioned to improve granularity; such a concept is not totally new for the LHCb collaboration, as LHCb SciFi trackers are opting for this option \cite{LHCb_SciFi}. Furthermore, the flat mirror of LHCb RICH-1, which reflects the photons to the sensor surface, has been considered to enter the acceptance and increase the curvature of the spherical mirrors. Such an optical system ensures improved emission point uncertainty as well as reduces occupancy and pixel error \cite{LHCb_Split_Optics}. However, this brings the flat mirror within the acceptance, and hence R\&D efforts towards reducing the material budget are assured.  

The gases with C$_n$F$_{n+2}$ structures are known as Saturated Fluro Carbons (SFC). Their non-conductivity and optical properties (large refractive index and limited chromaticity) are key elements for their use as Cherenkov radiators. Their non-conductivity, non-flammability, and radiation-resistance also make these gases excellent coolants. These gases, also currently in use by the LHCb RICH detectors, have high impact in the environment due to their high Global Warming Potential (GWP). The GWP of C$_4$F$_{10}$ is roughly 8500, similarly, the GWP of CF$_4$ is roughly 7000. In future, stricter policy on the usage of such gases imposes severe challenges in the operation of the detectors. LHCb primarily foresees the replacement of CF$_4$ with CO$_2$. However, this brings larger chromatic aberration and lower photon yield for RICH2, and the issue with RICH1 remains unsolved. Therefore, identification and validation of an alternative radiator gas are of extreme importance for LHCb. Newer spur-oxygenated fluoro-ketones, for example from the 3M NOVEC\textsuperscript{\textregistered} range, with C$_n$F$_{2n}$O structures, can offer similar performance but with very low, or zero GWP. Nevertheless, several aspects like flammability, toxicity, radiation hardness, non-conductivity, optical properties are of extreme importance and therefore extensive R\&D is ongoing in this direction \cite{Hallewell}. Such an effort bring synergy to the quest for a similar solution by the ePIC experiment and the dual RICH detector in the Future Circular Collider (FCC) \cite{Harnew_ECFA}. 
\subsection{TORCH}
Low momentum PID plays a critical role in reducing the systematic uncertainty by suppressing the background of key measurements related to CP violation. The PID can be enhanced, particularly for low momentum hadrons, by incorporating an innovative 30~m$^2$ area TOF detector in the LHCb setup to provide charged $\pi/K~(K/p)$ separation up to 10 (20)~GeV/c momentum and beyond. To obtain such performance, the time of propagation of Cherenkov light in a fused silica window is measured with sensors of high timing performance. The TORCH (Time of internally Reflected CHerenkov light) detector would require a 15~ps timing resolution on a track by track basis to achieve a $3 \sigma$ $\pi/K$ separation at 10~GeV/c. On one hand, such a detector would serve the role of the aerogel radiator of RICH1 that was envisaged in the original LHCb design; on the other hand, it would serve as a general purpose timing detector~\cite{LHCb_TORCH,LHCb_TORCH_Status}. It has been reported that TORCH can provide a 25 to 50\% improvement in the tagging efficiency of the initial state flavor tag, crucial for CP-violation measurements.

The TORCH aims to use planes of 1 cm thick fused silica as a source of prompt Cherenkov photons. It facilitates a modular design. Cherenkov photons travel to the periphery of the fused silica  plates by total internal reflection, where they are reflected by a cylindrical mirror surface of a quartz block. This surface focuses the photons onto a plane of MCP-PMTs where their positions and arrival times are measured. The expectation is that typically 30 photons will be detected; hence, the required ToF resolution dictates the timing of single photons to a precision of around 70~ps.
The MCP-PMTs have to withstand a high radiation environment beyond the current state of the art of 30~Ccm$^{-2}$; therefore, the granularity and occupancy per pixel are subject to optimization. In the baseline design the TORCH is operated with MCP-PMTs. The current generation of MCP-PMTs are known to have limitations in lifetime (due to ion feedback) and rate capability (from strip current), although improvements have been made, for example, through the use of ALD techniques. Improving the lifetime and rate capability of vacuum photon detectors is one of the goals of the CERN DRD4 collaboration. As an alternative, SiPMs are considered for the central regions of TORCH, where the occupancies are highest. SiPMs, however, have the drawback of large DCR, particularly when exposed to neutron fluences of 10$^{13}$ n$_{eq}$ /cm$^2$ or more. In order to mitigate the DCR, it may be necessary to cool the SiPMs. Such a solution may adapt a common approach like the LHCb scintillating fibre detectors, where SiPMs are considered as a technology of choice. Currently, the FastRICH has been considered as the baseline for readout. 

In 2018, a test beam was conducted, demonstrating that a 70~ps single photon time resolution is achievable \cite{LHCb_TORCH_BT18}. In 2025, a full-scale prototype was recently installed at the CERN PS beamline~\cite{LHCB_TORCH_BT25_Talk} and several critical aspects, such as mechanical design and optical gluing, have been validated. 
\begin{figure*}
    \centering
    \includegraphics[width=0.62\textwidth]{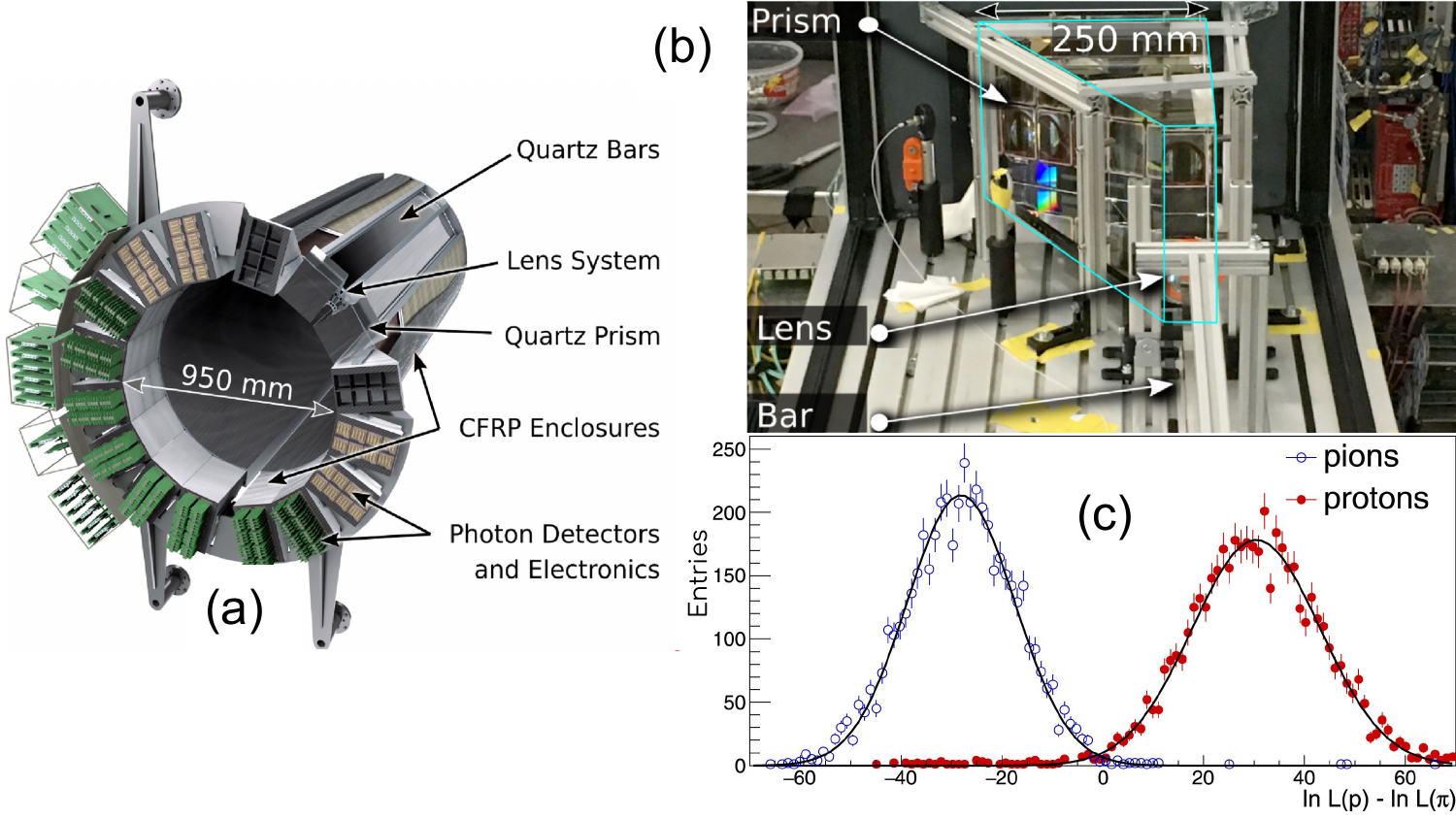}
    \vskip-1.2em
    \caption{PANDA barrel DIRC R\&D highlights: (a) realistic model of the PANDA barrel DIRC (b) the focusing components during beam test, (c) Separation of 7~GeV/c pion and proton samples during the beam test.}
    \label{fig:PANDA}
\end{figure*}
\section{DIRC detectors at PANDA:}
The PANDA experiment \cite{PANDA_TDR} at the international accelerator complex FAIR (Facility for Antiproton and Ion Research) in Darmstadt, Germany, will be using a high-quality antiproton beam with momenta in the range of 1.5 GeV/c to 15 GeV/c, stored in the HESR (High Energy Storage Ring), to explore fundamental questions of hadron physics in the charmed and multi-strange hadron sector and deliver decisive contributions to the open questions of Quantum ChromoDynamics (QCD). PID plays a central role in the physics program of PANDA. Cherenkov detectors, namely, DIRC detectors in the barrel region and in the end-cap region, are cornerstones of the PANDA physics program \cite{PANDA_bDIRC_TDR}. Originally inspired by the BaBar DIRC \cite{BaBar_DIRC} and key results from the R\&D activities of the focusing DIRC of the SuperB~\cite{SuperB_FDIRC}, the PANDA barrel DIRC has introduced novel ideas by replacing the water tank expansion volume with compact prisms, allowing it to operate near a magnetic field and insert focusing optics for further enhanced PID  (see Fig.~\ref{fig:PANDA} (a) for the main components of the PANDA barrel DIRC). The compact design of the detector requires the PANDA DIRC to place photo-sensors in a 2~T magnetic field and with a dense array of anode pixels. The single photon time resolution ($\sigma_t$) should be smaller than 100~ps to compensate for chromatic dispersion effects in the radiator bars and to measure the time-of-propagation of the photons. Moreover, the antiproton–proton average annihilation rate of 20~MHz in the PANDA experiment will lead to densities at the DIRC image planes of 200–800 kHz/cm$^2$ detected photons.

Microchannel Plates (MCPs) are chosen as the baseline photo-sensors for the PANDA barrel DIRC thanks to their several advantages; namely, they are commercially available as multi-anode devices and provide a good active area ratio while still being quite compact in size. MCP-PMTs exhibit high gain, high quantum efficiency, and good single photon sensitivity, excellent time resolution, much better magnetic field immunity compared to the classical MaPMTs, and low DCR; however they pose severe challenges in terms of aging \cite{ALICE_Lehmann_MCP}. In the last decade, MCP-PMTs were reported to show instability in the photocathode performance and a loss in quantum efficiency with an Integrated Anode Charge (IAC) of about 200~mC/cm$^2$, whereas the requirement of the PANDA experiment remains at a level of 5~C/cm$^2$. A mitigation strategy to lifetime degradation has been a major effort for the PANDA collaboration. It has been demonstrated that improvements in vacuum quality and sealing enhance the performance. However, an ultra-thin coating of resistive and/or secondary electron emissive (SEE) layer using ALD techniques applied to the surfaces of the MCP pores prevents the desorption of gaseous contaminants during the electron amplification process. An interesting side effect of this approach is the possibility of optimizing the MCP resistance and the SEE coefficient independently and hence increasing the rate capability and the gain of the MCP-PMT. A systematic and thorough characterization of several varieties of MCP-PMTs have demonstrated a `tremendously increased lifetime of MCP-PMTs' with ALD coating~\cite{ALICE_Lehmann_MCP}.

A dedicated beam test exercise has been performed at CERN, with Photonis MCP-PMT with extended lifetime. The results are in excellent agreement with the simulation studies performed with the barrel DIRC. Fig. \ref{fig:PANDA} (c) shows that the log likelihood difference of 7~GeV/c pions from protons is of the order of 5$\sigma$ separation. As direct $\pi$/K separation was not possible at T9 beam test facility; the observed separation between 7~GeV/c $\pi$/P was extrapolated to 3.5~GeV/c$~\pi$/K separation, thanks to the similar differences in the Cherenkov angle. The final design features narrow radiators made of synthetic fused silica, focusing optics with 3-layer spherical lenses and a compact prism-shaped expansion volume instrumented with MCP-PMTs~\cite{PANDA_bDIRC_BeamTest}.

\label{sec:PANDA}
\section{Use of PID at the ePIC experiment in EIC:}

\begin{figure*}
    \centering
    \includegraphics[width=0.62\textwidth]{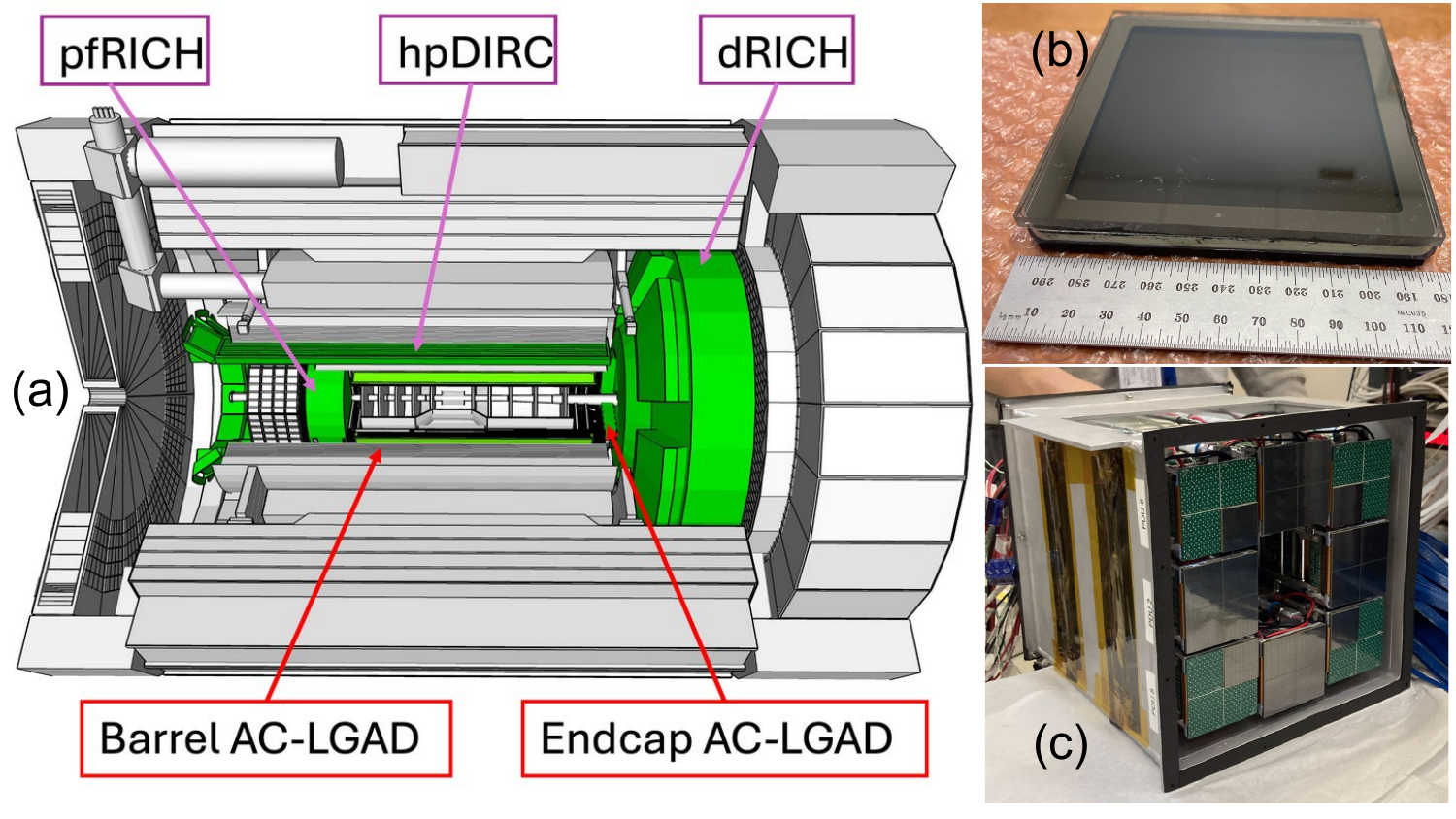}
    \vskip-1.2em
    \caption{PID in ePIC: (a) PID sub-detector systems, (b) HRPPD photo-sensor  (c) arrangements of SiPM sensors in beam test for dRICH}
    \label{fig:ePIC}
\end{figure*}

The ePIC experiment at the EIC is an approved project to be built at Brookhaven National Laboratory (BNL), USA. It aims to address key questions in the field of QCD and hadronic physics. The requirements for the EIC have been described in detail in the common effort by the physics community interested in the EIC physics, known as the Yellow Report \cite{EIC_YR}. In the Yellow Report, the role of the PID detectors in delivering the physics of the EIC has been considered fundamental. The extensive and challenging PID requirement, like in ePIC, cannot be delivered by a single PID technology: in fact, an ample phase-space domain has to be covered, including a wide momentum range. Therefore, several technologies have been identified to deliver the PID requirements altogether. The ePIC experiment employs three different Cherenkov-imaging-based technologies: a dual-radiator RICH in the forward direction (1.5~$\leq \eta \leq$~3.5), a proximity-focusing RICH in the backward direction ($-$3.5~$\leq \eta \leq$~$-$1.5), and a high-performance DIRC in the barrel ($-$1.5~$\leq \eta \leq$~1.5) region (Fig. \ref{fig:ePIC} (a)). The PID detectors have several challenges and R\&D activities related to each of these detectors primarily target solutions to these challenges, risk mitigation and optimization of the detector design \cite{EIC_Ullrich}. 

\subsection{The dual-radiator RICH:}
In the hadron-going direction, where most high-energy hadrons are produced, a dual-radiator Ring Imaging Cherenkov (dRICH) detector will be used to identify electrons, pions, kaons, and protons over a momentum range from a few hundred MeV/c up to 50~GeV/c and over a wide pseudorapidity range (1.5~$\leq \eta \leq$~3.5). To achieve this, a Cherenkov angle resolution of approximately 0.3–0.4 mrad on a track-by-track basis is required. The detector will also provide $e/\pi$ separation from a few hundred ~MeV/c up to 15~GeV/c. Six identical spherical mirrors will focus Cherenkov photons onto six spherical photo-sensor surfaces \cite{EIC_DRICH}. The photo-sensors are located in regions with magnetic fields of up to 1 T and at large angles with respect to the incident photon direction. Furthermore, due to the short radiator length (120~cm), the sensors are required to have high photon detection efficiency (PDE) to collect a sufficient number of photons.

Under these conditions, SiPMs emerge as the only cost-effective solution for covering large active areas, owing to their insensitivity to magnetic fields, high PDE, and excellent timing performance. However, SiPMs exhibit an intrinsic DCR and are susceptible to radiation damage. Radiation damage leads to an increase in dark current and therefore requires mitigation strategies, such as operating SiPMs at $-$30$~^{\circ}$C to reduce the DCR and applying periodical high-temperature annealing to recover from radiation damage. In addition, timing information can be exploited using high-precision TDC electronics with gating. Encouraging results on the performance of cooled SiPMs after irradiation and subsequent annealing cycles have been reported \cite{EIC_Preghenella, EIC_Radiation}.

To provide continuous performance over the full momentum range, the detector employs two radiators. A layer of hydrophobic aerogel produced by Aerogel Factory (Chiba, Japan) \cite{ALICE_ADACHI} covers the low-momentum region, providing at least $3\sigma$ $\pi/K$ separation up to 15 GeV/c, which corresponds to the Cherenkov threshold for kaons in the C$_2$F$_6$ gas used as the second radiator. Extensive studies have been carried out to optimize the optical parameters of the aerogel in order to extend the momentum reach while maintaining significant separation. These studies demonstrate that aerogel with a higher refractive index (1.026) provides better performance than the previously considered material with refractive index 1.02 \cite{EIC_aerogel}.

The dRICH will use C$_2$F$_6$ as the radiator gas for the first time in a RICH detector, enabled by its low chromatic dispersion. Studies have been performed to ensure that scintillation backgrounds remain sufficiently low and that the optical transparency of the radiator gas is high. In addition, a pressurized dRICH using argon as a radiator gas is being considered, with the pressure chosen to match the refractive index of C$_2$F$_6$, in anticipation of future policies on SFC gases. These R\&D activities involve interdisciplinary fields and may find applications in the FCC era. Dedicated R\&D on the gas system is a central aspect of the ePIC dRICH program and aligns with the scientific interests of DRD4 \cite{EIC_Gas}.

Several beam test studies have been performed with a prototype \cite{EIC_TB_Proto} to validate key features of the dRICH, including radiator optical quality and optical layout. The prototype, in its default configuration, uses Hamamatsu MaPMTs and MAROC chips \cite{MAROC}. The analysis shows that the single-photon resolution for both the aerogel and gas radiators is in excellent agreement with simulation studies. Furthermore, irradiated SiPMs with ALCOR chip–based readout \cite{EIC_ALCOR} have also been tested with this prototype (see Fig.~\ref{fig:ePIC} (c) for SiPM arrangements). A high number of photons were detected, with clear separation between gas and aerogel rings. The measured evolution of the ring radius with particle momentum enabled the identification of electrons, pions, kaons, and protons using aerogel photon information \cite{EIC_Rubini_SiPM}. Initial studies with a pressurized gas radiator have also been carried out to validate the number of detected photons and single-photon resolution. In the following years, a full scale dRICH prototype will be tested at CERN ~\cite{EIC_Marco_RICH25}. 

\subsection{A high-performance DIRC:}
A high-performance DIRC (hpDIRC) technology will provide PID in the ePIC barrel region by distinguishing pions and kaons up to 6 GeV/c with a $3\sigma$ separation. The hpDIRC is largely inspired by the PANDA DIRC design, as described in Section \ref{sec:PANDA}; the ePIC DIRC will reuse the BaBar DIRC bars \cite{BaBar_DIRC}. The central ideas behind achieving $\pi/K$ separation at such high momentum are the use of a novel focusing lens system to improve spatial resolution and the exploitation of the timing information of Cherenkov photons. Consequently, the hpDIRC is designed to use commercial MCP-PMTs. Furthermore, refurbishment of the bars, optical gluing of the bars to the expansion volume, and quality control are extremely delicate processes and are altogether progressing promisingly.

Dedicated simulation studies have demonstrated that, using realistic parameters of available commercial MCP-PMTs, the hpDIRC can provide the performance required for the ePIC physics program in the presence of a magnetic field. Currently, MCP-PMTs produced by PHOTEK (MAPMT253 series) are assumed as the baseline configuration. However, as a risk-mitigation strategy, DC-coupled HRPPD \cite{EIC_Lyashenko} photo-sensors are also being considered. HRPPD photo-sensors are chosen as the baseline photo-sensors for the pfRICH detector, highlighting the cross-cutting efforts among different subdetector systems within the ePIC community.

The PID performance required for the ePIC hpDIRC is technically challenging, and the targeted upper momentum limit is unprecedented for a detector of this type. Nevertheless, through the use of short radiator bars, optimized modular structures, a novel focusing lens system, and fast timing photo-sensors enabling reconstruction in both spatial and temporal domains, the hpDIRC aims to meet its PID requirements. Dedicated beam tests, accompanied by simulation studies and performed in collaboration with the PANDA barrel DIRC group, have shown promising results \cite{Kalicy:2024jme}.

\subsection{A Proximity-focusing RICH:}
In the electron-going direction, a proximity-focusing RICH (pfRICH) has been chosen as the baseline detector. The ePIC pfRICH is expected to provide at least 3$\sigma$ $\pi/K$ separation up to 7~GeV/c. Driven by the ePIC physics program, this region is subject to overlapping Cherenkov rings from outgoing jets. Therefore, the pfRICH aims to resolve these rings using not only spatial information but also the timing of Cherenkov photons. The pfRICH departs from the classical proximity-focusing approach by employing a large proximity gap to improve angular resolution ($\sim$~45~cm). To increase acceptance, conical mirrors are placed around the vessel wall and beam pipe. This, however, results in complex photon topologies on the sensor surface, particularly for photons emitted near the acceptance boundaries. To address the challenges of overlapping rings and ensure the required performance, the exploitation of timing information is central to pfRICH R\&D efforts.

Cherenkov photons in the pfRICH will be detected using HRPPDs produced by INCOM, using the same MCP-PMTs employed for LAPPDs \cite{EIC_LAPPD, EIC_Lyashenko}. Simulation studies have demonstrated that, with a nominal single–photoelectron time resolution of 50~ps, the pfRICH can achieve $\pi/K$ separation with high efficiency and low contamination. The photosensors also function as timing detectors. In a dedicated test beam at CERN, gen-II LAPPDs achieved 80~ps time resolution \cite{EIC_LAPPD_beamtest}. By exploiting the timing information of Cherenkov photons produced at the detector window by through-going charged particles, the HRPPD sensors can provide precise timing measurements. Thanks to their high quantum efficiency ($\sim$30\% \cite{EIC_Lyashenko}), the pfRICH will be capable of detecting an average of 12 photons per track.

The HRPPDs are subject to loss of gain and PDE in the presence of a magnetic field, particularly at large angles. However, a series of experiments with LAPPDs, a closely related technology produced by the same manufacturer, have demonstrated that both gain and PDE can be recovered by applying a higher bias voltage \cite{EIC_LAPPD_Mag}. Recently, HRPPDs have also been tested in magnetic fields at CERN and BNL. Preliminary results show that HRPPDs present no show-stoppers for pfRICH applications. For readout, a constant-fraction discriminator developed at Fermilab (FCFD) is considered as the baseline \cite{EIC_FCFD}. Within the EIC community, the AC-LGAD-based TOF in the barrel region will use a similar readout, ensuring a synergistic approach among different subdetectors.

In addition, degradation of the photocathode material under high radiation is a concern not only for the ePIC collaboration but for the broader Cherenkov-imaging community. Studies are being carried out at BNL, INFN Trieste, and JLab to understand photocathode aging under high illumination. Dedicated aging studies, supported by simulations of physics samples, have not indicated any significant performance degradation for the accumulated charge expected over the lifetime of the ePIC experiment. Dedicated R\&D activities for the pfRICH are ongoing on several fronts, including vessel preparation and in-house construction and characterization of the mirrors \cite{EIC_pfRICH_BPage}.

\section{Other relevant R\&D activities related to Cherenkov imaging:}
The R\&D activities are not restricted to the activities mentioned in the previous sections. Important efforts have been reported to be ongoing for many experiments worldwide. In the following part of this section, we discuss a few of them to understand the efforts. 

The Electron Ion Collider China (EIcC) will be designed with a similar philosophy but with substantially different kinematic coverage. Although the maximum momentum of outgoing hadrons is not comparable to that of the US EIC, almost hermetic PID coverage is critical. The baseline PID detector choices are similar to those of the ePIC experiment. Customized production of MCP-PMT, aerogel, SiPM and ASIC development are the  main motivations for the R\&D related to their PID subsystems. Further details can be found elsewhere \cite{EIC_China}.

The Super Tau Charm Factory (STCF) also requires efficient PID. The barrel PID system is comprised of a proximity focusing RICH, and liquid C$_6$F$_{14}$ has been considered as a radiator~\cite{STCF_CDR}. Hybrid Micro Pattern Gaseous Detector (MPGD) \cite{MPGD} based photo-sensors based on CsI coated THick Gas Electron Multiplier (THGEM) \cite{THGEM} and Micromegas \cite{Micromegas} have been chosen as the default option. The set-up is supposed to provide $\pi/K$ separation up to 2~GeV/c. Custom built radiator purification, THGEM and Micromegas production, and design of the readout chain are the main directions of their R\&D; recently, SiPMs are also being considered as an alternative solution \cite{STCF_MPGD}. 

The J-PARC hadron facility also aims for a dual radiator RICH with aerogel (n=1.04) and C$_4$F$_{10}$ gas radiator \cite{MARQ}. The required single photon angular resolution is around 10~mrad. Currently, the hydrophobic aerogel produced by the Aerogel Factory has been chosen as the baseline, and Hamamatsu MPPC S13360-6075CS has been selected as the photo-sensors. To increase the photo-sensitive area, aluminized Mylar foils are considered as light concentrators. Beam test exercises with prototype and ongoing R\&D activities are focused on optimizing the light guide, SiPM cooling, and aerogel characteristics \cite{MARQ}. On the other hand, the Belle-II experiment is working on finding an optimal photon detector to replace the HAPD photo-sensors~\cite{Belle2_HAPD}. After 2030 the production of HAPDs are expected to discontinue. Multi-Pixel Photon Counter (MPPC) arrays and LAPPDs are under consideration. Dedicated R\&D efforts related to the sensors and ASIC design are ongoing \cite{Belle2_LAPPD}.
\begin{table*}[h]
\centering
\small
\begin{tabular}{|l|l|}
\hline
\textbf{Synergy}    & \textbf{Experiments (PID subsystem)}   \\ \hline
Aerogel             & ePIC dRICH, ALICE3 bRICH, possibly FCC \\ \hline
SFC gas alternative & ePIC dRICH, LHCb, J-PARC, FCC            \\ \hline
SiPM (MPPC array) &
  ePIC dRICH, ALICE3, LHCb, J-PARC, possibly FCC, possibly BelleII, CBM upgrade \\ \hline
LAPPD/HRPPD &
  \begin{tabular}[c]{@{}l@{}}ePIC pfRICH, possibly ALICE3 fRICH, possibly LHCb, alternative solution \\ for ePIC hpDIRC\end{tabular} \\ \hline
\begin{tabular}[c]{@{}l@{}}Commercial MCP\\ (PHOTEK)\end{tabular} &
  LHCb TORCH, PANDA barrel DIRC,  ePIC hpDIRC (baseline) \\ \hline
Readouts &
  \begin{tabular}[c]{@{}l@{}}\textit{ALCOR}: ePIC dRICH, ALICE bRICH; \textit{FCFD}: ePIC pfRICH, ePIC hpDIRC\\ \textit{FastRICH}: LHCb RICH and TORCH\end{tabular} \\ \hline
\end{tabular}
\vskip -0.5 em
\caption{Summary of possible technology synergies}
\label{tab:synergy}
\end{table*}

At CERN, the ALADDIN experiment \cite{ALADDIN_2} aims to achieve a precise measurement of the electric dipole moment  of charmed baryons ($\Lambda_{c}^+$ and $\Xi_{c}^+$). Suppression of decay products of $D^+$ and $D_S^{+}$ is important to ensure efficient $\Lambda_{c}^+$ reconstruction. To reduce background, ALADDIN plans to use a focusing RICH detector for separating protons from kaons and pions in a range from a few hundred GeV up to the TeV range. This exceptional momentum range requires a large radiator. Currently, the ALADDIN RICH is 750~cm long. The choice of photo-sensor is critical, simulation studies and R\&D activities are ongoing within the DRD4 collaboration. Both SiPM and commercial MCP-PMTs are being considered; ALCOR chips are taken as the baseline readout. Depending on the sensor choice, the radiator gas may be He or Ne. He gas may be incompatible with MCP-PMTs due to degassing. The final design choice is extensively studied with R\&D efforts, test beams, and dedicated simulation studies \cite{ALADDIN_1}. Looking further ahead, the CLD experiment at the FCC-ee aims to use an Array of RICH Cells (ARC), where each cell is a  dual radiator RICH \cite{ARC}. The aerogel radiator serves both as a radiator and a thermal insulator between the cooled SiPM and C$_4$F$_{10}$ gas. The challenge remains to detect enough photons with an extremely short radiator length ($\sim 20$ cm). While the detector is currently validating its proof of principle, it opens several R\&D opportunities synergistic to the LHCb and ePIC experiments. Dedicated simulation studies are ongoing.  

Recent tests with the prototype RICH of the CBM experiment demonstrated the compatibility of the DiRICH with the novel DAQ concept developed. In the baseline design, the MaPMT (H12700) is used. A future upgrade with SiPM photo-sensors has been considered and subject to ongoing R\&D activities \cite{CBM_Pauly_RICH25,CBM_RICH}. 

\section{Conclusion and Acknowledgments}
In conclusion, we can safely say that ongoing synergistic R\&D activities (see the summary in Table \ref{tab:synergy}) related to the imaging application of Cherenkov light for PID in future particle and nuclear physics experiments are broad and encouraging. Several global trends in Cherenkov imaging technology are clearly emerging. First, there is an increasing demand for high-timing-performance sensors to enable time imaging and potential TOF applications using prompt Cherenkov photons. This demand has driven a global preference for SiPMs or commercial MCP-PMTs, either as baseline sensors or as potential future upgrades. Both types of sensors offer advantages as well as intrinsic limitations. A significant portion of the R\&D effort focuses on selecting the appropriate sensor based on physics requirements and mitigating their limitations to achieve optimized performance.

The choice of SFC radiator gas for focusing RICH detectors remains a challenge. Several experiments, including LHCb and ePIC, retain SFC gases as their baseline choice, but they are actively exploring alternatives such as Novac gases or pressurized gases with matched refractive indices. A coordinated, synergistic effort to identify alternative radiator solutions will undoubtedly benefit the development of future RICH detectors.  

The author is extremely grateful to Silvia Gambetta, Giacomo Volpe, Jianbei Liu, and Silvia Dalla Torre for their valuable insights and contributions to this work.

\footnotesize
\bibliography{refs}

@misc{DRD4,
    key = "{DRD4 collaboration: Development of Photon Detectors and Particle Identification Techniques}",
    note ={},
    url= "{https://drd4.web.cern.ch/}" 
}

@article{ALICE,
    author = {Acharya, S. and others },
    title = "{The ALICE experiment: a journey through QCD}",
    journal ="{ Eur. Phys. J. C 84, 813 (2024)}",
    year = 2024,
    doi = "{https://doi.org/10.1140/epjc/s10052-024-12935-y}",
    url={https://link.springer.com/article/10.1140/epjc/s10052-024-12935-y}
}

@article{ALICE_LOI,
    collaboration = "ALICE",
    title = "{Letter of intent for ALICE 3: A next-generation heavy-ion experiment at the LHC}",
    eprint = "2211.02491",
    archivePrefix = "arXiv",
    primaryClass = "physics.ins-det",
    reportNumber = "CERN-LHCC-2022-009, LHCC-I-038",
    month = "11",
    year = "2022",
    url="https://inspirehep.net/literature/2176715"
}

@article{ALICE_Scope,
    author = {Acharya, S. and others},
    title = "{Scoping document for ALICE 3: ALICE phase IIb upgrade for the LHC Long Shutdown 4}",
    journal = "{CERN-LHCC-2025-002, LHCC-G-185}",
    year = 2025,
    url = "https://cds.cern.ch/record/2925455?ln=en"
}

@article{ALICE_Lehmann_MCP,
    author = "Lehmann, A. and others",
    title = "{Latest Technological Advances with MCP-PMTs}",
    doi = "10.1088/1742-6596/2374/1/012128",
    journal = "J. Phys. Conf. Ser.",
    volume = "2374",
    number = "1",
    pages = "012128",
    url ="https://iopscience.iop.org/article/10.1088/1742-6596/2374/1/012128"
}

@inproceedings{ALICE_Nicassio,
    author = "Nicassio, Nicola and others",
    title = "{A combined SiPM-based TOF+RICH detector for future high-energy physics experiments}",
    doi = "10.1109/iwasi58316.2023.10164558",
    month = "6",
    year = "2023",
    url = "https://ieeexplore.ieee.org/document/10164558"
}

@article{ALICE_ADACHI,
    title = "{Status of high-quality silica aerogel radiators}",
    journal = {NIMA},
    volume = {952},
    pages = {161919},
    year = {2020},
    doi = {https://doi.org/10.1016/j.nima.2019.02.046},
    url = {https://www.sciencedirect.com/science/article/pii/S0168900219302347},
    author = {Ichiro Adachi},
}

@article{ALICE_TB,
    author = "Altamura, A. R. and others",
    title = "{Beam test studies for a SiPM-based RICH detector prototype for the future ALICE~3 experiment}",
    eprint = "2601.12511",
    archivePrefix = "arXiv",
    primaryClass = "physics.ins-det",
    doi = "10.1140/epjc/s10052-025-14287-7",
    journal = "Eur. Phys. J. C",
    volume = "85",
    number = "5",
    pages = "578",
    year = "2025",
    url="https://link.springer.com/article/10.1140/epjc/s10052-025-14287-7"
}

@article{ALICE_Rad1,
    author = "Altamura, Anna Rita and others",
    title = "{Radiation damage on SiPMs for space applications}",
    eprint = "2112.08089",
    archivePrefix = "arXiv",
    primaryClass = "physics.ins-det",
    doi = "10.1016/j.nima.2022.167488",
    journal = "Nucl. Instrum. Meth. A",
    volume = "1045",
    pages = "167488",
    year = "2023",
    url ="https://www.sciencedirect.com/science/article/pii/S016890022200780X"
}

@article{ALICE_SiPM,
    author = "Calvi, M. and Carniti, P. and Gotti, C. and Matteuzzi, C. and Pessina, G.",
    title = "{Single photon detection with SiPMs irradiated up to 10$^{14}$ cm$^{-2}$ 1-MeV-equivalent neutron fluence}",
    eprint = "1805.07154",
    archivePrefix = "arXiv",
    primaryClass = "physics.ins-det",
    doi = "10.1016/j.nima.2019.01.013",
    journal = "Nucl. Instrum. Meth. A",
    volume = "922",
    pages = "243--249",
    year = "2019"
}

@techreport{LHCb_RICH1,
      author        = "Brook, N and others",
      collaboration = "LHCbRICH",
      title         = "{LHCb RICH1 Engineering Design Review Report}",
      institution   = "CERN",
      reportNumber  = "LHCb-2004-121, CERN-LHCb-2004-121",
      address       = "Geneva",
      year          = "2005",
      url           = "https://cds.cern.ch/record/897981",
}

@techreport{LHCb_RICH2,
      author        = "Adinolfi, M and others",
      title         = "{LHCb RICH 2 engineering design review report}",
      institution   = "CERN",
      reportNumber  = "LHCb-2002-009",
      address       = "Geneva",
      year          = "2002",
      url           = "https://cds.cern.ch/record/691478",
      note          = "revised version number 1 submitted on 2002-05-21 14:24:22",
}

@book{LHCB_RICH_TDR,
      author        = "Amato, S and others",
      collaboration = "LHCb",
      title         = "{LHCb RICH}",
      publisher     = "CERN",
      address       = "Geneva",
      series        = "Technical design report. LHCb",
      year          = "2000",
      url           = "https://cds.cern.ch/"
}

@techreport{LHCb_Upgrade_TDR,
      author        = "Lindner, Rolf and others",
      title         = "{LHCb Particle Identification Enhancement Technical Design
                       Report}",
      institution   = "CERN",
      reportNumber  = "CERN-LHCC-2023-005, LHCB-TDR-024",
      address       = "Geneva",
      year          = "2023",
      url           = "https://cds.cern.ch/record/2866493",
      doi           = "10.17181/CERN.LAZM.F5OH",
}

@article{LHCb_performance,
      author        = "Calabrese, R. and others",
      title         = "{Performance of the LHCb RICH detectors during LHC Run 2}",
      archivePrefix = "arXiv",
      eprint        = "2205.13400",
      reportNumber  = "LHCb-DP-2021-004",
      journal       = "JINST",
      volume        = "17",
      number        = "07",
      pages         = "P07013",
      year          = "2022",
      url           = "https://cds.cern.ch/record/2811056",
      doi           = "10.1088/1748-0221/17/07/P07013",
}

@article{LHCb_CLARO,
    doi = {10.1088/1748-0221/12/08/P08019},
    url = {https://doi.org/10.1088/1748-0221/12/08/P08019},
    year = {2017},
    month = {aug},
    publisher = {},
    volume = {12},
    number = {08},
    pages = {P08019},
    author = {Baszczyk, M. and others},
    title = {{{CLARO}}: an {{ASIC}} for high rate single photon counting with multi-anode photomultipliers},
    journal = {JINST},
}

@article{LHCb_FastIC,
    doi = {10.1088/1748-0221/17/05/C05027},
    url = {https://doi.org/10.1088/1748-0221/17/05/C05027},
    year = {2022},
    month = {may},
    publisher = {IOP Publishing},
    volume = {17},
    number = {05},
    pages = {C05027},
    author = {Gómez, S and others},
    title = {{{FastIC}}: a fast integrated circuit for the readout of high performance detectors},
    journal = {JINST},
}

@article{LHCb_FastRICH,
    title = "{The FastRICH ASIC for the {{LHCb}} RICH enhancements}",
    journal = {NIMA},
    volume = {1067},
    pages = {169664},
    year = {2024},
    issn = {0168-9002},
    doi = {https://doi.org/10.1016/j.nima.2024.169664},
    url = {https://www.sciencedirect.com/science/article/pii/S0168900224005904},
    author = {F. Keizer},
}

@misc{lpGBT,
    key = "{lpGBT Project Collaboration; Low Power GigaBite Transceiver (webpage)}",
    note={},
    year={2018},
    url ={https://lpgbt-fpga.web.cern.ch/doc/html/index.html},
}

@conference{VLPlus,
    author = {Vasey, F and others},
    booktitle={},
    title = "{ Versatile Link PLUS Information to Users and Q\&A}",
    url = {https://indico.cern.ch/event/799025/sessions/317612/attachments/1901284/3138633/VLplus_Vasey_TWEPP_03Sep2019.pdf},
    year = {2019},
}

@article{LHCb_SciFi,
    title = "{Microlens-enhanced SiPMs for the {{LHC}}b {{SciFi}} tracker Upgrade II: Update and recent results}",
    journal = {NIMA},
    volume = {1080},
    pages = {170607},
    year = {2025},
    issn = {0168-9002},
    doi = {https://doi.org/10.1016/j.nima.2025.170607},
    url = {https://www.sciencedirect.com/science/article/pii/S0168900225004085},
    author = "Ronchetti, F and others"
}

@conference{LHCb_Split_Optics,
    author = {Cardinale, R and others } ,
    booktitle = {DRD4 WG4 "Software" Meeting},
    title ={Flexible and customizable ray-tracing for optical design with OpticaEM} ,
    year = {March 2025},
}

@article{Hallewell,
    author = {Hallewell, G},
    title ="{The “green” use of fluorocarbons in Cherenkov detectors and silicon tracker cooling systems: challenges and opportunities in an unfolding era of alternatives}",
    journal = {Eur. Phys. J. Plus (2023) 138: 1141},
    year ={ 19 December 2023 },
    url={https://doi.org/10.1140/epjp/s13360-023-04703-w}
}

@article{Harnew_ECFA,
    title = "{The {{ECFA}} Roadmap process for particle identification and photon detector R\&D}",
    journal = {NIMA},
    volume = {1057},
    pages = {168768},
    year = {2023},
    issn = {0168-9002},
    doi = {https://doi.org/10.1016/j.nima.2023.168768},
    url = {https://www.sciencedirect.com/science/article/pii/S0168900223007593},
    author = {N. Harnew}
}

@article{LHCb_TORCH,
    title = "{TORCH: Time of flight identification with Cherenkov radiation}",
    journal = "{NIMA}",
    volume = {639},
    number = {1},
    pages = {173-176},
    year = {2011},
    doi = {https://doi.org/10.1016/j.nima.2010.09.021},
    url = {https://www.sciencedirect.com/science/article/pii/S0168900210020140},
    author = {M.J. Charles and R. Forty},
}

@article{LHCb_TORCH_Status,
    title = "{Status of the TORCH time-of-flight project}",
    journal = "{NIMA}",
    volume = {952},
    pages = {161692},
    year = {2020},
    note = {10th International Workshop on Ring Imaging Cherenkov Detectors (RICH 2018)},
    issn = {0168-9002},
    doi = {https://doi.org/10.1016/j.nima.2018.12.007},
    url = {https://www.sciencedirect.com/science/article/pii/S016890021831800X},
    author = "Harnew, N and others",
}

@article{LHCb_TORCH_BT18,
    title = "{Performance of a prototype TORCH time-of-flight detector}",
    journal = {NIMA},
    volume = {1050 },
    pages = {168181},
    year = {2023},
    doi = {https://doi.org/10.1016/j.nima.2023.168181},
    url = {https://www.sciencedirect.com/science/article/pii/S0168900223001717},
    author = {Bhasin, S. and others},
}

@conference{LHCB_TORCH_BT25_Talk,
    author = "Lehuraux, M and others",
    booktitle = "{TORCH detector concept and design; RICH 2025}",
    title = "{TORCH detector concept and design}",
    year = "{2025}"
}

@article{PANDA_TDR,
    author = "Kotulla, M and others",
    title = "{Technical Progress Report for: PANDA}",
    journal ="{FAIR-ESAC/Pbar/Technical Progress Report}" ,
    url = "https://panda.gsi.de/oldwww/archive/public/panda_tpr.pdf"
}

@article{PANDA_bDIRC_TDR,
    author = "Singh, B and others",
    title = "{Technical design report for the $\bar P$ANDA Barrel DIRC detector}",
    doi = "10.1088/1361-6471/aade3d",
    journal = "J. Phys. G",
    volume = "46",
    number = "4",
    pages = "045001",
    year = "2019"
}

@article{BaBar_DIRC,
    author ="Adam, I and others" ,
    title = "{The DIRC particle identification system for the BaBaR experiment}",
    journal ="{NIMA}",
    volume= {538},
    pages = {281},
    year = {2005},
    doi= {doi:10.1016/j.nima.2004.08.129},
}

@article{SuperB_FDIRC,
    author ="Roberts, D. A. and others" ,
    title = "{Results from the FDIRC prototype}",
    journal ="{NIMA}" ,
    volume={766},
    pages={114-117},
    year = {2014},
    doi= {https://doi.org/10.1016/j.nima.2014.04.069}
}

@article{PANDA_bDIRC_BeamTest,
    author = "Dzhygadlo, D and others",
    title = "{The PANDA Barrel DIRC}",
    journal = "{NIMA}",
    volume="{1055}",
    pages="{168480}",
    year = 2023
}

@article{EIC_YR,
    title = "Science Requirements and Detector Concepts for the Electron-Ion Collider: EIC Yellow Report",
    journal = {Nuclear Physics A},
    volume = {1026},
    pages = {122447},
    year = {2022},
    issn = {0375-9474},
    doi = {https://doi.org/10.1016/j.nuclphysa.2022.122447},
    url = {https://www.sciencedirect.com/science/article/pii/S0375947422000677},
    author = {Abdul Khalek, R and others}
}

@article{EIC_Ullrich,
	author = {Ullrich, Thomas},
	title = "{Requirements and R\&D for detectors at the future Electron–Ion Collider}",
    journal ={NIMA},
	year = {2022},
	volume = {1039},
    pages="167041",
	doi = {10.1016/j.nima.2022.167041}
}

@article{EIC_DRICH,
	author = {Del Dotto, A. and others},
	title = "{Design and R\&D of RICH detectors for EIC experiments}",
	journal ="NIMA",	
	year = {2017},
	volume = {876},
	pages = {237 – 240},
	doi = {10.1016/j.nima.2017.03.032},	
}

@conference{EIC_Marco_RICH25,
    author = "{M. Contalbrigo}",
    title = "{The ePIC Dual-Radiator RICH Detector; RICH 2025 conference, Mainz}",
    year = 2025,
}

@article{EIC_Preghenella,
    title = "{A SiPM-based optical readout system for the EIC dual-radiator RICH}",
    journal = {NIMA},
    volume = {1046},
    pages = {167661},
    year = {2023},
    issn = {0168-9002},
    doi = {https://doi.org/10.1016/j.nima.2022.167661},
    url = {https://www.sciencedirect.com/science/article/pii/S0168900222009536},
    author = {Preghenella, R. and others},
}

@article{EIC_ALCOR,
    title = "{ALCOR: A mixed-signal ASIC for the dRICH detector of the ePIC experiment at the EIC}",
    journal = {NIMA},
    volume = {1069},
    pages = {169817},
    year = {2024},
    issn = {0168-9002},
    doi = {https://doi.org/10.1016/j.nima.2024.169817},
    url = {https://www.sciencedirect.com/science/article/pii/S0168900224007435},
    author = {Cossio, F and others},
}

@article{EIC_Rubini_SiPM,
    title = "{The SiPM readout plane for the ePIC-dRICH detector at the EIC: Overview and beam test results}",
    journal = {NIMA},
    volume = {1082},
    pages = {170890},
    year = {2026},
    issn = {0168-9002},
    doi = {https://doi.org/10.1016/j.nima.2025.170890},
    url = {https://www.sciencedirect.com/science/article/pii/S0168900225006928},
    author = {Rubini, N and others},
}

@article{EIC_Radiation,
    title="{Study of radiation effects on SiPM for an optical readout system for the EIC dual-radiator RICH}",
    author={Preghenella, R and others},
    journal={NIMA},
    volume={1056},
    pages={168578},
    year={2023},
    publisher={Elsevier}
}

@article{MAROC,
    author = "Blin, Sylvie and Barrillon, Pierre and de La Taille, Christophe",
    editor = "Keyser, Ron",
    title = "{MAROC, a generic photomultiplier readout chip}",
    doi = "10.1109/NSSMIC.2010.5874062",
    journal = "IEEE Nucl. Sci. Symp. Conf. Rec.",
    volume = "2010",
    pages = "1690--1693",
    year = "2010"
}

@article{EIC_TB_Proto,
title = "{Prototype of a dual-radiator RICH detector for the Electron–Ion Collider}",
journal = {NIMA},
volume = {1058},
pages = {168834},
year = {2024},
issn = {0168-9002},
doi = {https://doi.org/10.1016/j.nima.2023.168834},
url = {https://www.sciencedirect.com/science/article/pii/S0168900223008252},
author = {Vallarino, S. and others}
}

@conference{EIC_Gas,
    author = "Tessarotto, F and others",
    title = "{The ePIC dRICH radiator gas, RICH 2025, Mainz}",
    year = {2025}
}

@article{EIC_aerogel,
  author = "Altamura, Anna Rita",
  title = "{Silica aerogel characterization for the ePIC dRICH detector}",
  doi = "10.22323/1.476.1116",
  journal = "PoS",
  year = 2024,
  volume = "ICHEP2024",
  pages = "1116"
}

@article{EIC_Lyashenko,
  title="{HRPPD photosensors for RICH detectors with a high resolution timing capability}",
  author={Lyashenko, A and and others},
  journal={NIMA},
  pages={170964},
  year={2025},
  publisher={Elsevier}
}

@article{EIC_LAPPD,
    title = "{Performance of Large Area Picosecond Photo-Detectors (LAPPDTM)}",
    journal = {NIMA},
    volume = {958},
    pages = {162834},
    year = {2020},
    note = {Proceedings of the Vienna Conference on Instrumentation 2019},
    issn = {0168-9002},
    doi = {https://doi.org/10.1016/j.nima.2019.162834},
    url = {https://www.sciencedirect.com/science/article/pii/S0168900219312690},
    author = {Lyashenko, A.V. and others},
}

@article{EIC_LAPPD_beamtest,
    author = "Bhattacharya, D. S. and others",
    title = "{Characterization of LAPPD timing at CERN PS testbeam}",
    eprint = "2309.15011",
    archivePrefix = "arXiv",
    primaryClass = "physics.ins-det",
    doi = "10.1016/j.nima.2023.168937",
    journal = "NIMA",
    volume = "1058",
    pages = "168937",
    year = "2024"
}

@article{EIC_LAPPD_Mag,
    author = "Agarwala, Jinky and others",
    title = "{Performance of an LAPPD in magnetic fields}",
    doi = "10.1016/j.nima.2024.170122",
    journal = "NIMA",
    volume = "1072",
    pages = "170122",
    year = "2025"
}

@conference{EIC_pfRICH_BPage,
    author = "Page, B and others",
    title = "{A Proximity Focusing RICH Detector for the ePIC Experiment at the EIC, Mainz, RICH 2025}",
    year = 2025
}

@article{Kalicy:2024jme,
    author = "Kalicy, G.",
    title = "{The high-performance DIRC for the ePIC detector at the EIC}",
    doi = "10.1016/j.nima.2024.169168",
    journal = "NIMA",
    volume = "1062",
    pages = "169168",
    year = "2024"
}

@article{EIC_FCFD,
    title = "{Design and performance of the Fermilab Constant Fraction Discriminator ASIC}",
    journal = {NIMA},
    volume = {1056},
    pages = {168655},
    year = {2023},
    issn = {0168-9002},
    doi = {https://doi.org/10.1016/j.nima.2023.168655},
    url = {https://www.sciencedirect.com/science/article/pii/S0168900223006459},
    author = {Xie, S. and others},
}

@article{EIC_China,
    author ={Anderle, D.P. and others} ,
    title = "{Electron-ion collider in China}",
    journal = "{Frontiers of Physics}",
    volume ={16(6)},
    pages={ 64701},
    year = 2021
}

@article{STCF_CDR,
    author = "Achasov, M and others",
    title = "{STCF conceptual design report (Volume 1): Physics \& detector}",
    journal = "{Frontiers of Physics}",
    volume= {19},
    doi ="{https://doi.org/10.1007/s11467-023-1333-z}",
    year = 2024,
}

@techreport{CLD,
      author        = "Bacchetta, N. and others",
      collaboration = "CERNLinearColliderDetector",
      title         = "{CLD - A Detector Concept for the FCC-ee}",
      institution   = "CERN",
      archivePrefix = "arXiv",
      eprint        = "1911.12230",
      reportNumber  = "LCD-Note-2019-001",
      address       = "Geneva",
      year          = "2019",
      url           = "https://cds.cern.ch/record/2697140",
      note          = "75 pages, 67 figures",
}

@article{BelleII,
  title="{Belle II technical design report}",
  author={Abe, T. and others},
  journal={arXiv preprint arXiv:1011.0352},
  year={2010}
}

@article{STCF_MPGD,
    doi = {10.1088/1748-0221/20/06/C06012},
    url = {https://doi.org/10.1088/1748-0221/20/06/C06012},
    year = {2025},
    month = {jun},
    publisher = {IOP Publishing},
    volume = {20},
    number = {06},
    pages = {C06012},
    author = "Wang, A and others",
    title = "{Study of MPGD based photodetectors for the STCF-RICH system}",
    journal = "{JINST}",
}

@article{MARQ,
    author = "Suzuki, K and others",
    title = "{Development of a cost-effective RICH detector for MARQ spectrometer using SiPM technology}",
    journal = "{JINST}",
    volume="{20}",
    pages= {C06070},
    url="{https://doi.org/10.1088/1748-0221/20/06/C06070}",
    year = {2025}
}

@conference{ALADDIN_1,
    author = "Spadaro, E and others",
    title = "{A RICH detector for the ALADDIN experiment, RICH 2025, Mainz}",
    year = 2025
}

@techreport{ALADDIN_2,
      author        = "Akiba, K and others",
      title         = "{ALADDIN: An Lhc Apparatus for Direct Dipole moments INvestigation}",
      institution   = "CERN",
      reportNumber  = "CERN-LHCC-2024-011, LHCC-I-041",
      address       = "Geneva",
      year          = "2024",
      url           = "https://cds.cern.ch/record/2905467",
      doi           = "10.17181/CERN.2G4V.0YAO",
}

@article{CBM,
  author       = {Ablyazimov, T. and Abuhoza, A. and Adak, R. P. and others},
  title        = {Challenges in QCD matter physics -- The scientific programme of the Compressed Baryonic Matter experiment at FAIR},
  journal      = {European Physical Journal A},
  year         = {2017},
  volume       = {53},
  pages        = {60},
  doi          = {10.1140/epja/i2017-12248-y},
  url          = {https://doi.org/10.1140/epja/i2017-12248-y}
}

@article{CBM_RICH,
title = "{Qualification of DIRICH readout chain}",
journal = {NIMA},
volume = {1056},
pages = {168570},
year = {2023},
issn = {0168-9002},
doi = {https://doi.org/10.1016/j.nima.2023.168570},
url = {https://www.sciencedirect.com/science/article/pii/S0168900223005600},
author = {Becker, M and others },
}

@conference{CBM_Pauly_RICH25,
    author = "Pauly, C and others",
    title = "{Status of the CBM RICH detector -towards first beam in 2028, RICH 2025, Mainz}",
    year = 2025
}

@manual{ARC,
    title        = "\emph{Simulation and performance study of the ARC concept: a compact RICH for future collider experiments}",
    author       = "Cardinale, R and others",
    year         = 2024,
    journal      = "{CERN Repository:}",
    doi          = {10.17181/6entj-pmm10},
    url          = {https://doi.org/10.17181/6entj-pmm10},
}

@article{Belle2_LAPPD,
title = "{Development and performance evaluation of readout systems for Belle II ARICH upgrade}",
journal = {NIMA},
volume = {1080},
pages = {170695},
year = {2025},
issn = {0168-9002},
doi = {https://doi.org/10.1016/j.nima.2025.170695},
url = {https://www.sciencedirect.com/science/article/pii/S0168900225004966},
author = {Kurokawa, S. and others},
}

@article{Belle2_HAPD,
title = "{A 144-channel HAPD for the Aerogel RICH at Belle II}",
journal = {NIMA},
volume = {766},
pages = {145-147},
year = {2014},
issn = {0168-9002},
doi = {https://doi.org/10.1016/j.nima.2014.05.060},
url = {https://www.sciencedirect.com/science/article/pii/S0168900214006007},
author = {Korpar, S. and others},
}

@book{MPGD,
  title="{Micro-pattern gaseous detectors: Principles of operation and applications}",
  author={Sauli, Fabio},
  year={2020},
  publisher={World Scientific}
}

@article{THGEM,
title = "{The Thick Gas Electron Multiplier and its derivatives: Physics, technologies and applications}",
journal = {Progress in Particle and Nuclear Physics},
volume = {130},
pages = {104029},
year = {2023},
issn = {0146-6410},
doi = {https://doi.org/10.1016/j.ppnp.2023.104029},
url = {https://www.sciencedirect.com/science/article/pii/S0146641023000108},
author = {Bressler, S. and others},
}

@article{Micromegas,
title = {MICROMEGAS: a high-granularity position-sensitive gaseous detector for high particle-flux environments},
journal = {NIMA},
volume = {376},
number = {1},
pages = {29-35},
year = {1996},
issn = {0168-9002},
doi = {https://doi.org/10.1016/0168-9002(96)00175-1},
url = {https://www.sciencedirect.com/science/article/pii/0168900296001751},
author = {Y. Giomataris and Ph. Rebourgeard and J.P. Robert and G. Charpak},
}
\end{document}